# A 2000-year record of arsenic variability from a South Pole ice core

Elena V. Korotkikh[1,2*], Paul A. Mayewski[1,2], Andrei V. Kurbatov[1,2], Kirk Maasch[1,2], Jefferson C. Simões[3,1], Michael J. Handley[1], Sharon B. Sneed[1], Daniel A. Dixon[1], Mariusz Potocki[1,2]

[1] Climate Change Institute, University of Maine, Orono, Maine 04469-5790, USA

[2] School of Earth and Climate Sciences, University of Maine, Orono, ME, USA

[3] Centro Polar e Climático, Instituto de Geociências, Universidade Federal do Rio Grande do Sul, Av. Bento Gonçalves, 9500 C.P. 15001 91501-970 - Porto Alegre, RS - Brazil

Corresponding author: Elena V. Korotkikh (elena.korotkikh@maine.edu)

**Abstract.** Using a South Pole ice core covering the last ~2050 years we present a high-resolution (4-27 samples/year), continuous record of natural and anthropogenic arsenic (As) deposition. Our results show that volcanic emissions (notably from Mt. Erebus) are a significant source of As for South Pole, and human activities potentially contributed to As deposition as early as 225 C.E. The most significant anthropogenic source As enrichment in the record, starting in 1975 C.E., is linked to increased copper production in Chile and at least partially to coal combustion from throughout the Southern Hemisphere. East and West Antarctic ice core delivered deposition differences are a result of differences in the atmospheric circulation patterns that transport As to these regions.

# 1. Introduction

Arsenic is one of the most toxic elements in the environment and can be found in low concentrations throughout the geosphere. It can enter the environment through both natural processes and anthropogenic activities [*Chilvers and Peterson*, 1987]. Several studies report increases in As concentrations during the late 20th century in soil, water, and air worldwide. In many cases elevated As is attributed primarily to human activities [*Bhattacharya et al.*, 2007; *Reimann et al.*, 2009]. The residence time of As in the atmosphere, between 7 and 10 days [*Matschullat*, 2000], allows As to be transported over relatively long distances. Ice core records provide an excellent opportunity to reconstruct the history of As concentrations in the atmosphere in remote regions. Atmospheric models and volcanic cloud transport data show that air masses from South America, Australia and southern Africa can reach the Antarctic continent in ~10 days [*Li et al.*, 2010; *Moxey*, 2010; *Albani et al.*, 2012; *Dixon et al.*, 2012; *Sudarchikova et al.*, 2015], therefore the Antarctic continent is potentially capturing anthropogenic pollution from Southern Hemisphere countries. Elevated concentrations of several trace elements (Pb, Cu, Cr, Zn, Ag, U and Bi) are documented in Antarctic ice core records and are attributed to emissions from smelting and mining operations, combustion of fossil fuel and use of leaded gasoline in Australia and South America [*Planchon et al.*, 2002; *Hur et al.*, 2007; *Potocki et al.*, 2016]. Concentrations of As have been reported in a few Antarctic ice core records that cover the last 45-125 years, and it is suggested that enrichment during the second half of the 20$^{th}$ century is attributed to increased copper smelting in Chile [*Hong et al.*, 2012; *Rong et al.*, 2016; *Schwanck et al.*, 2016]. Enrichment in As during the last 3000 years due to metallurgical activities in the Northern Hemisphere is reported in an ice core record from Devon Ice Cap in the Arctic [*Krachler et al.*, 2009]. Therefore, it is possible that human activities also affected atmospheric As deposition prior to the 20$^{th}$ century in the Southern Hemisphere.

Here we present a continuous high-resolution record of atmospheric As concentrations from a South Pole ice core for the period from 60 B.C.E. to 1999 C.E. The goal of this paper is to investigate As sources impacting Antarctica and to discuss variability related to natural versus anthropogenic sources.

## 2. Methodology

### 2.1. Ice core collection and chemical analysis

The South Pole SPRESSO (South Pole Remote Earth Science and Seismological Observatory) ice core was collected as part of the International Trans Antarctic Science Expedition (ITASE) (site 02-6) at 89.93°S, 144.39°W, at altitude of 2808 m (Figure 1) [*Mayewski et al.*, 2006; *Korotkikh et al.*, 2014]. No drilling fluid was used in the recovery of the core thus avoiding a major potential source of contamination.

The top 200 meters of the South Pole ice core was sampled using the Climate Change Institute (CCI) continuous melting system (see methodology in *Osterberg et al.*, 2006). Samples from the inner part of the core were collected for ICP-SFMS (Inductively Coupled Plasma Sector Field Mass Spectrometry) and IC (Ion Chromatography) analysis at an average sample resolution of ~1cm. ICP-SFMS samples were acidified to 1% with double-distilled $HNO_3$ and allowed to react with acid for at least 2 months before analysis. Every sample from sections 0.88-59.4 m and 148.9-161 m depth (sample resolution 4-27 samples/year), and every tenth sample from the rest of the core (sample resolution 1-2 samples/year) were analyzed for major and trace elements using the Thermo Electron Element2 ICP-SFMS. All samples were analyzed for their major anion ($Cl^-$, $NO_3^-$, $SO_4^{2-}$) content using a Dionex DX-500 ion chromatograph. This study is focuses on changes in As concentration in the South Pole ice core chemistry record. As concentrations were measured at low resolution (LR) mode. The LR mode is preferable for analyzing elements with very low concentrations, because of its higher sensitivity. Comparison with HR mode data shows that the LR data are not affected by potential As interferences (Figure S1). Detection limit for As is 0.13 ng/L. Detection limits for other elements used in this study are shown in Table S1.

### 2.2. Dating of the ice core

The South Pole ice core record was annually dated by counting seasonal peaks from several elements and calibrated using major historical volcanic eruptions as age markers (Figure S2). The top 60 meters of the ice core was dated using seasonal peaks in Na, Sr, S, $SO_4^{2-}$ and $Cl^-$ concentrations. The 60 to 200-meter depth section was dated using $Cl^-$ and $SO_4^{2-}$ seasonal peaks.

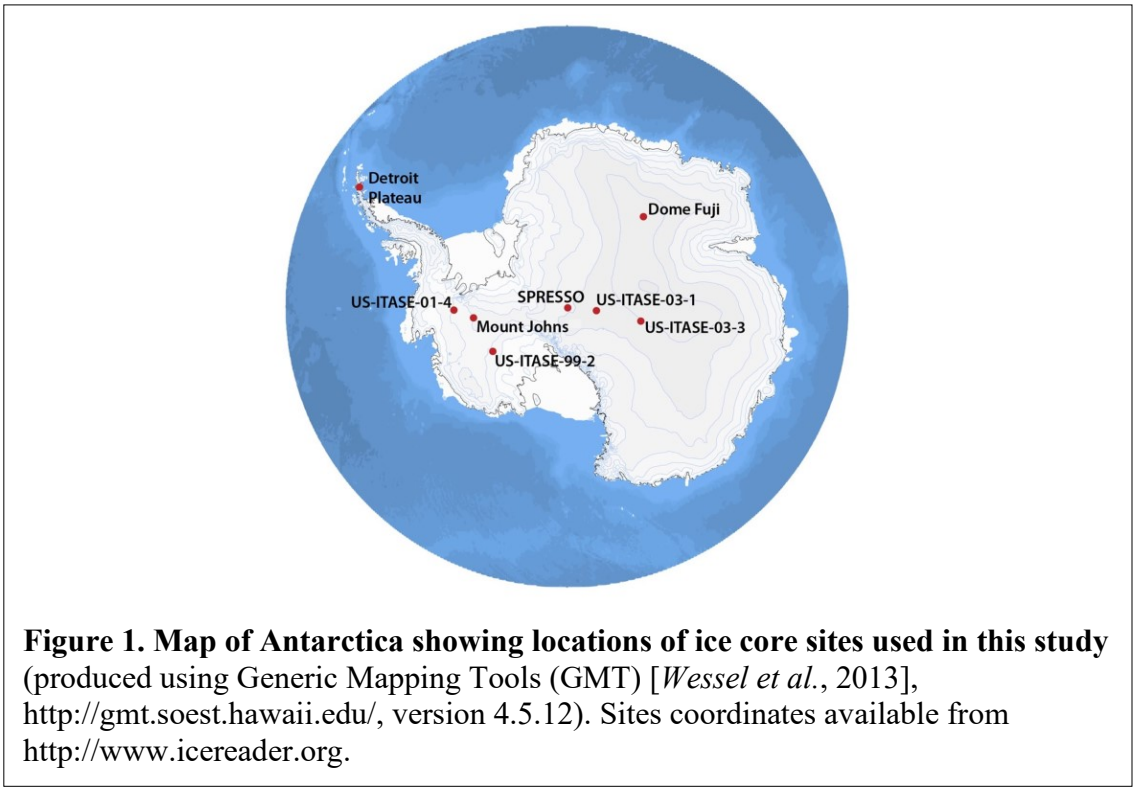

**Figure 1. Map of Antarctica showing locations of ice core sites used in this study** (produced using Generic Mapping Tools (GMT) [*Wessel et al.*, 2013], http://gmt.soest.hawaii.edu/, version 4.5.12). Sites coordinates available from http://www.icereader.org.

Once we developed a preliminary timescale and established volcanic signal markers (identified by large peaks in S and $SO_4^{2-}$ concentrations), we compared the pattern of volcanic signals in our record with the volcanic record in the WAIS divide ice core record [*Sigl et al.*, 2013]. The WAIS divide ice core has a higher accumulation rate, therefore, the annual signal is better preserved and there is less of a chance to have missing layers. We set the age of the volcanic events using the WAIS timescale and adjusted our record accordingly. On the basis of our dating, the South Pole record covers the period from -66 to 1999 C.E. Estimated dating errors are: for the period 1963-1999 C.E. - ±1 year; 1815-1963 C.E. - ±3 years; 1458-1815 C.E. - ±11 years; 1257-1458 C.E. - ±12 years; 232-1257 C.E. - ±24 years. Because of the absence of any established volcanic time markers in the deeper section of the core, we were not able to estimate dating error below 232 C.E.

## 3. Results and discussion

### 3.1. As concentrations for the period from 60 B.C.E. to 1999 C.E.

Figure 2 shows the high-resolution (1-27 samples/year) record of As variability developed from the South Pole ice core. Concentrations of As in our record are low and vary from 0.13 to 26.15 ng/L, with an average concentration of 1.8 ng/L. Several periods of increased As concentration were observed in the South Pole record: ~225-325 C.E., ~645-985 C.E., ~1260-1400 C.E., and ~1730-1999 C.E. (Table 1). The most significant increase in As concentrations is observed after 1975 C.E., when concentrations average 6.6 ng/L, which is four times higher than during the period 60 B.C.E. to 1974 C.E.

As flux (concentration*annual accumulation rate) is also shown on Figure 2. Both As flux and As concentration demonstrate similar variability, indicating that the changes in As concentrations in South Pole records are not associated with changes in snow accumulation rate.

Table 1. Mean concentrations of South Pole As during different periods, and estimated contributions from natural sources and their percentage relative to the measured concentrations.

| | As | Sea spray | | Volcanic | | Soil dust | |
|---|---|---|---|---|---|---|---|
| **Periods with elevated As concentration:** | ng/L | ng/L | % | ng/L | % | ng/L | % |
| 225-325 C.E. | 2.5 | 0.004 | 0.2 | 0.6 | 29.2 | 0.1 | 6.4 |
| 645-985 C.E. | 2.5 | 0.006 | 0.2 | 0.6 | 29.2 | 0.1 | 6.2 |
| 1260-1400 C.E. | 2.7 | 0.006 | 0.3 | 0.5 | 27.3 | 0.1 | 5.3 |
| 1730-1974 C.E. | 2.4 | 0.005 | 0.2 | 0.4 | 21.7 | 0.1 | 3.1 |
| 1975-1999 C.E. | 6.5 | 0.003 | 0.0 | 0.3 | 5.5 | 0.1 | 1.3 |
| **Periods with low As concentration:** | | | | | | | |
| 60 B.C.E.-224 C.E. | 1.0 | 0.005 | 0.5 | 0.5 | 80.1 | 0.1 | 12.7 |
| 326-644 C.E. | 0.9 | 0.003 | 0.4 | 0.6 | 92.2 | 0.1 | 8.0 |
| 986-1259 C.E. | 1.4 | 0.006 | 0.7 | 0.5 | 79.2 | 0.1 | 7.6 |
| 1401-1729 C.E. | 1.1 | 0.004 | 0.5 | 0.4 | 59.7 | 0.0 | 5.0 |

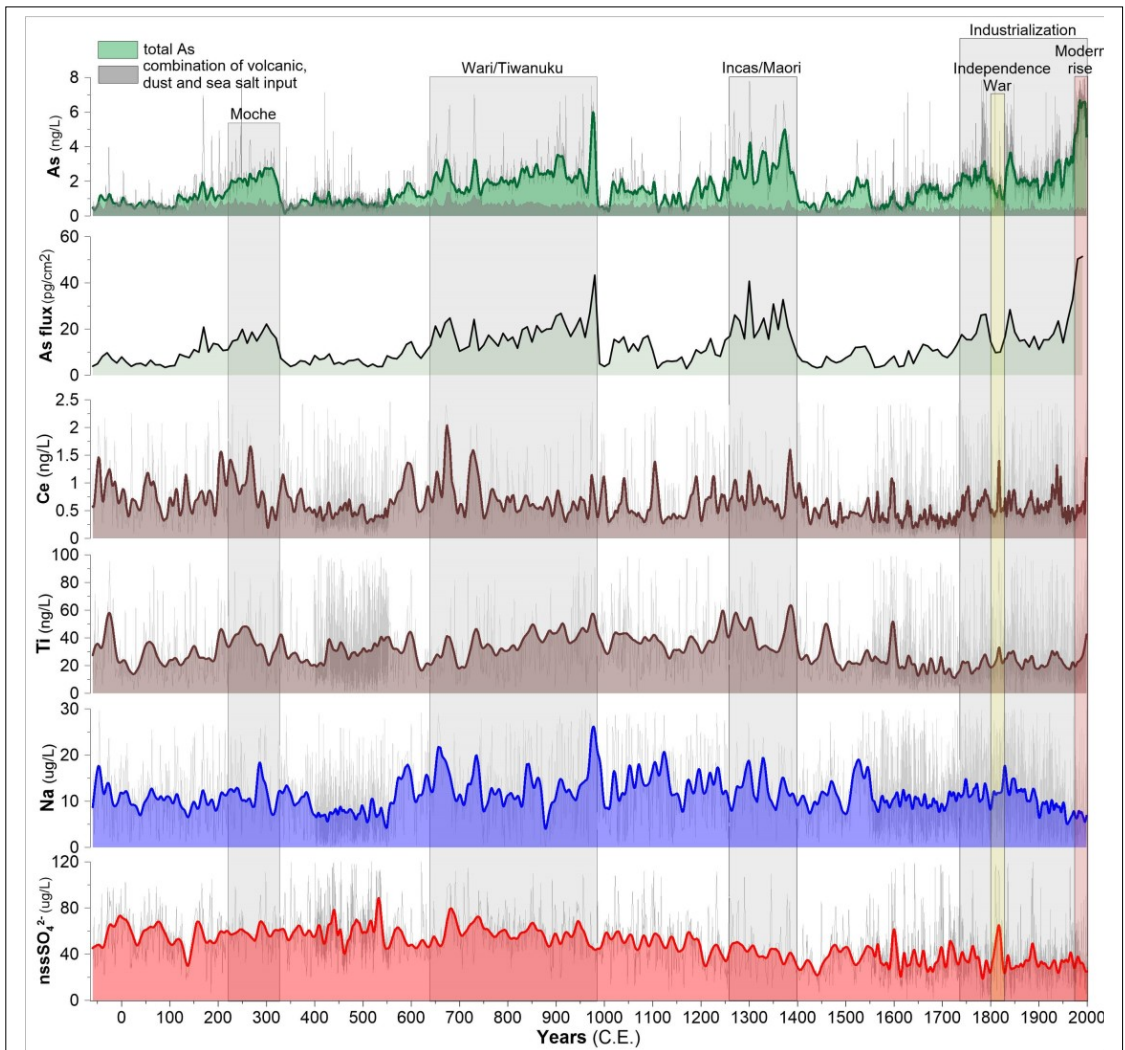

**Figure 2. South Pole records from top: As (ng/L), As flux (pg/cm$^2$), Ce (ng/L), Ti (ng/L), Na (ug/L), and nssSO$_4^{2-}$ (ug/L) for the period from 60 B.C.E. to 1999 C.E.** The light grey color lines show raw values. The colored lines represent background levels estimated using a robust spline smoothing function. Note that some raw values exceed the vertical scale. Flux is shown as 10-year resampled data. Colored horizontal bands define periods of changes in As concentrations potentially associated with anthropogenic activities.

**3.2. Estimation of contribution from natural sources.**

Arsenic can enter the atmosphere from several natural sources including crustal and soil dust, sea salt aerosols, wild forest fires, volcanic and biogenic emissions [*Nriagu*, 1989]. Previous studies suggest that explosive volcanic eruptions and quiescently degassing volcanoes are the primary natural source for As [*Chilvers and Peterson*, 1987; *Nriagu*, 1989].

To evaluate the natural contribution to South Pole As, we calculated soil dust, sea salt spray and volcanic emission input. The sea salt spray maximum contribution is estimated using the following equation:

$$As_{sea\ salt} = Na_{sample} * ([As/Na]_{ocean\ water}),$$

where ocean elemental abundances are from Lide [2005].

The As contribution from soil dust is calculated using the following equation:

$$As_{soil\ dust} = X_{sample} * ([As/X]_{soils}),$$

where X is a reference element and concentrations of elements in soils are from *Kabata-Pendias* [2011]. We estimated a mean soil dust contribution using Ti, Pr, La and Ce as reference elements.

Our source partitioning calculations show that the sea spray contribution is negligible (<1%) throughout the whole record (Table 1). This is similar to previous estimations and other observations in Antarctic records [*Chilvers and Peterson*, 1987; *Hong et al.*, 2012; *Schwanck et al.*, 2016]. Contributions from soil dust are also small and on average account for ~5% of total As (Table 1).

Contributions from quiescent volcanic degassing are calculated using $SO_4^{2-}$ concentrations in South Pole samples and mean values of As/S ratios in volcanic emissions. We assume that 2% of Antarctic $SO_4^{2-}$ comes from degassing volcanoes [*Cosme et al.*, 2005]. We recognize that estimation of the volcanic contribution is complex, because As/S values in published compilations vary largely. The volcanic contribution estimates based on As/S ratios from different studies are shown in Figure S3. The highest As/S values are reported from Mount Erebus, an active volcano located in East Antarctica, that potentially could be a source of impurities at the South Pole and is suggested to have a regional impact over Antarctica [*Zreda-Gostynska et al.*, 1997]. Our calculations, using a maximum As/S concentration ratio value of 0.001364 from Mt

Erebus emissions [*Zreda-Gostynska et al.*, 1997], show that regional volcanism could contribute on average ~50% to the As budget at South Pole. The regional volcanic contribution is especially significant during periods of low As deposition at South Pole where it can account for ~ 80% of the total annual As budget. The volcanic contribution is relatively lower during periods of increased As concentrations (~20%) (Table 1), suggesting that during these times other sources are more important to the As budget.

Another important natural source for As is biologically mediated marine and terrestrial volatilization [*Nriagu*, 1989; *Zhang et al.*, 2013]. Increases in organoarsenic compounds in surface waters have been associated with primary productivity as a result of methylation reactions by phytoplankton [*Zhang et al.*, 2013]. However, the lack of an As correlation (r=-0.1, p<0.0001) with non-sea-salt$SO_4^{2-}$ (nss$SO_4^{2-}$) in our record (Figure 2), an indicator of marine biogenic productivity in Antarctica [*Cosme et al.*, 2005], suggests that marine biogenic input to South Pole As is not significant. Previous studies estimate that low temperature volatilization from soils could also contribute to the As budget [*Chilvers and Peterson*, 1987; *Nriagu*, 1989], however, published studies do not allow us to make any quantitative estimations of the terrestrial biogenic contribution for Antarctica. It is possible, though, that terrestrial biogenic emission could be responsible for the elevated As concentration.

We estimate that during periods of low As concentration a significant component comes from volcanic emissions and a relatively small fraction from soil dust (Table1). Periods of low As concentrations usually correspond to intervals with decreased dust (Ti, La, Ce, Pr) and sea salt (Na) concentrations (Fig. 2). Decreases in dust elements and Na are linked to changes in atmospheric circulation, and therefore a decrease in As transport from more distant sources at these times, leaving the As budget at these times likely dominated by a regional volcanic source such as Mt. Erebus.

In contrast during periods with high As deposition (Table1), the volcanic contribution is less significant because As is supplied to South Pole from other sources. Intervals with increased As concentration typically coincide with an increase in Na and elements affiliated with dust (Ti, La, Ce, Pr) (Fig. 2) indicating enhanced aerosol loading of air transported to the South Pole. At these times it is likely that enhanced atmospheric circulation results in the transport of As from more distant sources, possibly also carrying biogenic As from Southern Hemisphere land masses. Intensified atmospheric circulation could, therefore, also transport anthropogenic source As from Southern Hemisphere countries.

**3.3. Anthropogenic emissions**. Arsenic can enter the atmosphere from anthropogenic activities such as: non-ferrous metal smelting and mining, fossil fuel combustion, pesticides, herbicides, wood preservatives, burning of pasture land, and glass manufacture [*Chilvers and Peterson*, 1987; *Matschullat*, 2000]. Non-ferrous metal production, especially copper smelting, is the most significant source for As and is estimated to account for ~70% of worldwide anthropogenic As in the late 1990s [*Pacyna*, 1987]. Copper and copper alloys ores have been smelted in South America for more than 3,000 years [*Cooke et al.*, 2008b; *Cortés and Scattolin*, 2017] and we suggest that these metallurgical activities could have contaminated the atmosphere as far away as Antarctica with As.

Figure 2 shows several intervals with elevated As concentrations, that we suggest based on the foregoing, could be related to human activities. Increase in As concentrations between 225-325 C.E. could be caused by metallurgical activities of the Mocher civilization in northern Peru [*Hörz and Kallfass*, 2000; *Eichler et al.*, 2017]. At the same time we observe an increase in dust concentrations indicating enhanced atmospheric circulation, thus As from more distant sources, such as northern Peru, could be transported to the South Pole site.

An increase in As concentrations between ~645-985 C.E. coincides with a period of active arsenic bronze metallurgy by the Wari and Tiwanaku empires in southern South America [*Cooke et al.*, 2008a]. Decreases in As concentrations around 985 C.E. coincide with the collapse of both empires (~1000 C.E.), suggesting that ancient native American metallurgical activity in South America potentially affected As deposition at the South Pole.

The next period of elevated As concentration (Figure 2) starting ~1260 C.E., corresponds to the rise of the Inca civilization. However, As concentrations decline ~1400 C.E. and remain low from 1400 to ~1725 C.E., despite an active Inca Empire and following widespread of the early Colonial metallurgy. Lower levels of As concentrations might be explained by two factors: a shift in atmospheric transport to South Pole, as evidenced by relatively lower dust loading during this period (Figure 2); and a decrease in copper smelting and intensification of silver metallurgy during Colonial times.

Increased As concentration between ~1250-1400 C.E. also coincides with a period of rapid deforestation and associated burning in New Zealand following the Maori arrival [*McWethy et al.*, 2009]. High severity fires during the "Initial Burning Period" resulted in a rapid loss of nearly half of the native forest. Such a large burning event would promote release of trace elements into the atmosphere and it is possible that increased As concentrations at South Pole at this time are related to the New Zealand forest fires.

Increases in As background concentration level since ~1730 correspond to the beginning of the industrial revolution. Numerous copper, gold and silver mines were opened in Chile during the 1700's, and the copper industry continued to expand during the 18th and 19$^{th}$ centuries [*Conrad J. Bahre*, 1979]. Decrease in As between ~1800-1830 C.E. is believed to be related to stagnation of the mining industry during the South American War of Independence, after which As increased significantly during the period ~1830-1850 C.E. This time interval coincides with the introduction of fuel-fired reverberatory furnaces for copper smelting in Chile and the subsequent increase in smelting activities [*Conrad J. Bahre*, 1979]. At the same time (~1840 C.E.) southern Australia experienced a mining boom following the discovery of silver-lead and copper ores [*Australian Mining History Association*, 2011], which could also contribute to South Pole As enrichment.

**3.4. Recent increase.** The most striking feature in the South Pole record is the As rise since 1975 C.E. As concentration increases 4.2 times, reaching a maximum in concentration of 26.15 ng/L, the highest value during the last 2050 years (Table 1, Figure 2). Dust elements and Na concentrations decrease during this time (Figure 2), indicating that the As increase is not related to the intensification of atmospheric circulation. Volcanic, sea salt and soil dust sources only account for a maximum of ~3% of the total As at this time, clearly indicating that the recent As enrichment is anthropogenic in origin. Previous studies show that air traffic and activities at research stations can contaminate the local Antarctic environment near these stations [*Wolff and Cachier*, 1998; *Mazzera et al.*, 2001; *Korotkikh et al.*, 2014] However, elevated levels of As in recent decades are evident in several other Antarctic records located far from stations (Figure 3), demonstrating that the South Pole ice core As increase is not solely related to local station contamination, but is more likely caused by hemispheric scale input integrated from different emission sources.

As mentioned above, Cu smelting is the biggest source of anthropogenic As [*Nriagu and Pacyna*, 1988; *Pacyna and Pacyna*, 2001]. Global modeling studies suggest that during the period 1999-2000 C.E. emissions from South America contributed up to 90% of the As deposition in Antarctica [*Wai et al.*, 2016].

Chile is the largest Cu producer in the world with an increasing trend in production since the early 19$^{th}$ century and a rapid rise since the 1970s [*British Geological Survey*, 2015] (Figure 3). We observe a significant correlation (r=0.83, p<0.0001) between As concentrations in the South Pole record and Chilean copper production for the period between 1847-1999 CE. This correlation suggests that Cu smelting in Chile is a likely cause of the As enrichment in the South Pole ice core record. We note a decrease in As concentration in early 1998 A.D, when As levels drop to an average of ~3.8 ng/L. The As level is still higher than during the pre-industrial period, however, it is significantly lower compared to 1970-1997 C.E. values. Other Antarctic records also show a decrease in As concentrations in the late 1990s (Figure 3). This decrease is coincident with the introduction of environmental regulations to limit airborne As emissions in Chile in 1990s. The first set of regulations were established in 1991, and further regulation was passed in 1999 [*Caldentey and Mondschein*, 2003]. The decline in As concentrations at the South Pole and other Antarctic records suggests that environmental regulations in Chile and consequent reduction of As emission from Cu smelters [*Caldentey and Mondschein*, 2003] produced successful outcome.

The second to largest anthropogenic source for atmospheric As after Cu smelting is coal combustion, which can account for up to 22% of anthropogenic As [*Nriagu and Pacyna*, 1988; *Pacyna and Pacyna*, 2001]. The consumption of fossil fuels in Southern Hemisphere countries has continuously increased during recent decades. South Africa, the biggest coal consumer in the Southern Hemisphere, burned ~1930 million tonnes of oil equivalent (Mtoe) of coal during 1965-2000 C.E., followed by Australia, which consumed ~950 Mtoe [*BP Statistical Review of World Energy*, 2015].

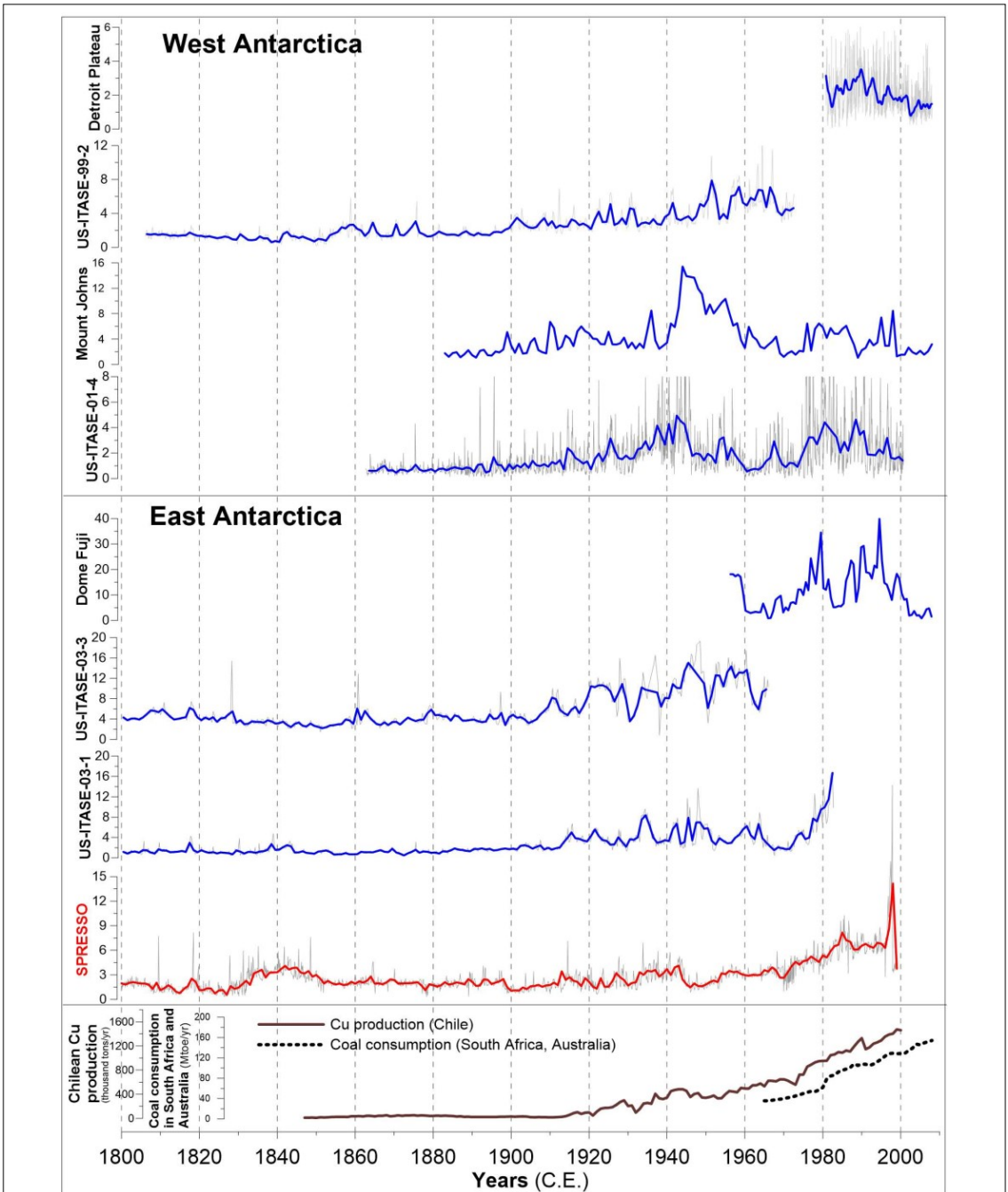

**Figure 3. Comparison of South Pole As concentrations with other Antarctic ice core records.** As records from the following sites are shown: Detroit Plateau [*Potocki et al.*, 2016], US-ITASE-99-2 [*Mayewski and Dixon*, 2013], US-Mount Jones [*Schwanck et al.*, 2016], ITASE-01-4 [*Mayewski and Dixon*, 2013], Dome Fuji [*Hong et al.*, 2012], US-ITASE-03-3 [*Mayewski and Dixon*, 2013], US-ITASE-03-1 [*Mayewski and Dixon*, 2013], and SPRESSO As concentrations (ng/L). The grey color lines show raw As data. The colored lines represent background levels estimated using a robust spline smoothing function. Chilean copper smelter production [*BritishGeologicalSurvey*, 2015], and Australian and South African coal consumption [*BP Statistical Review of World Energy*, 2015] are also shown**.**

We found high positive correlations between South Pole As concentrations and coal consumption in South Africa of r=0.6 (p<0.0001) and As and coal consumption in Australia of r=0.7 (p<0.0001) (Figure 3). This indicates that coal consumption is at least partially responsible for the enrichment in As at the South Pole during recent decades.

**3.5. Differences between East and West Antarctic As deposition.** Comparison of Antarctic As ice core records is difficult, since some of the records do not cover the period since 1975 C.E. or only show limited time intervals. Figure 3 reveals a difference between East and West Antarctic As records. East Antarctic records show a rapid increase in As concentration since ~1975 C.E. West Antarctic As records show a different pattern with two periods of elevated As concentrations: ~1935-1960 C.E. and ~1975-1999 C.E. The increase during the most recent period is not as significant as in the East Antarctic ice core records. While East Antarctic As records show a similar increasing trend to Chilean copper production, West Antarctic As concentrations do not reveal this association. The difference between As trends could be due to the different atmospheric circulation patterns impacting these regions. West Antarctica is strongly influenced by lower atmospheric transport, enhanced by cyclonic activity. More elevated inland East Antarctic sites are more influenced by mid and upper-troposphere air masses [*Sudarchikova et al.*, 2015].

A major low pressure system in the area of the Amundsen/Ross Seas, the Amundsen Sea Low (ASL), has a significant effect on climatic conditions in West Antarctica [*Raphael et al.*, 2016]. The strength and position of the ASL are affected by the Southern Annual Mode (SAM), a primary mode that controls extratropical Southern Hemisphere climate variability. Since the late 1950s, the SAM has been shifting to a more positive mode, with the major shift occurring in the mid-1970s [*Thompson and Wallace*, 2000; *Marshall*, 2003]. The shift to a positive SAM has resulted in intensification of westerly flow around Antarctica and the contraction of the polar vortex [*Mayewski et al.*, 2013]. While intensification of the westerlies leads to more efficient poleward transport to East Antarctica, it might have the opposite effect on West Antarctic sites. Stronger westerly flow has caused the ASL to move eastward closer to the Antarctic Peninsula [*Cullather et al.*, 1996; *Turner et al.*, 2013]. When the ASL is located farther eastward, one consequence is low net precipitation in West Antarctica [*Kreutz et al.*, 2000]. This might explain the lack of increase in As concentrations in West Antarctic sites since 1975 C.E.

Another factor that could affect As deposition in Antarctica is the different dominant dust source areas. Previous studies show that South American sources dominate dust deposition in East Antarctica, and that Australia is a more important source for West Antarctica [*Li et al.*, 2008; *Albani et al.*, 2012]. Thus, As emitted from copper smelting activities in Chile is more efficiently transported by westerly flow to downwind East Antarctic sites.

**Summary.** This study presents a high-resolution record of As variability for the period from 60 B.C.E. to 1999 C.E. The South Pole As record shows high variability with several periods of increased concentrations. Most of the As is attributed to volcanic emissions from Mount Erebus during periods of low As deposition. Intervals with increased As concentrations could be related to increased anthropogenic emissions in the Southern Hemisphere countries and partially to the changes in atmospheric transport. Our record suggests that anthropogenic activities in the Southern Hemisphere might have contaminated the natural As budget during the last two millennia: notably during 645-985 C.E. due to the mining and smelting activities by Wari and Tiwanaku empires and during 1260-1400 C.E. due to activities carried out by the Incas and possibly Maori. Elevated As concentrations since ~1730 C.E. are most likely related to anthropogenic activities in South America and Australia. The most significant increase in As occurs after 1975 C.E. and is attributed to increased copper smelting in Chile and partially to coal combustion in the Southern Hemisphere. A similar rise is observed in other East Antarctic sites, but not in West Antarctica. These differences in As deposition is attributed to the different atmospheric pathways and changes in atmospheric circulation since the mid-1970s. More specifically, the southward shift in the polar vortex since the 1970s and the consequent increase in the strength of the westerlies.

**Acknowledgments.**

This research was supported by NSF grants PRL-1042883, 0439589, 0636506, 0829227, 1203640. We greatly acknowledge the support of the US Antarctic Program, the 109th New York Air National Guard, Ice Coring and Drilling Services, Raytheon Polar Services Company, the National Ice Core Laboratory, Kenn Borek Air Ltd., and all of our US ITASE field colleagues.

**Supporting Information**

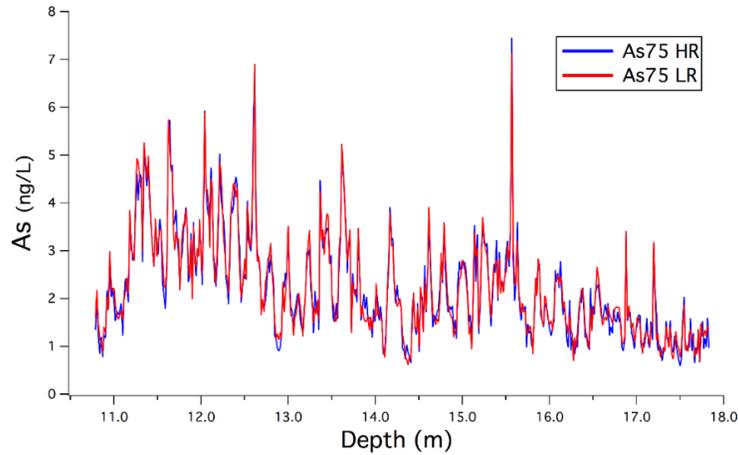

**Figure S1.** Comparison of As concentrations analyzed in low (LR) and high resolution (HR) ICP-SFMS modes.

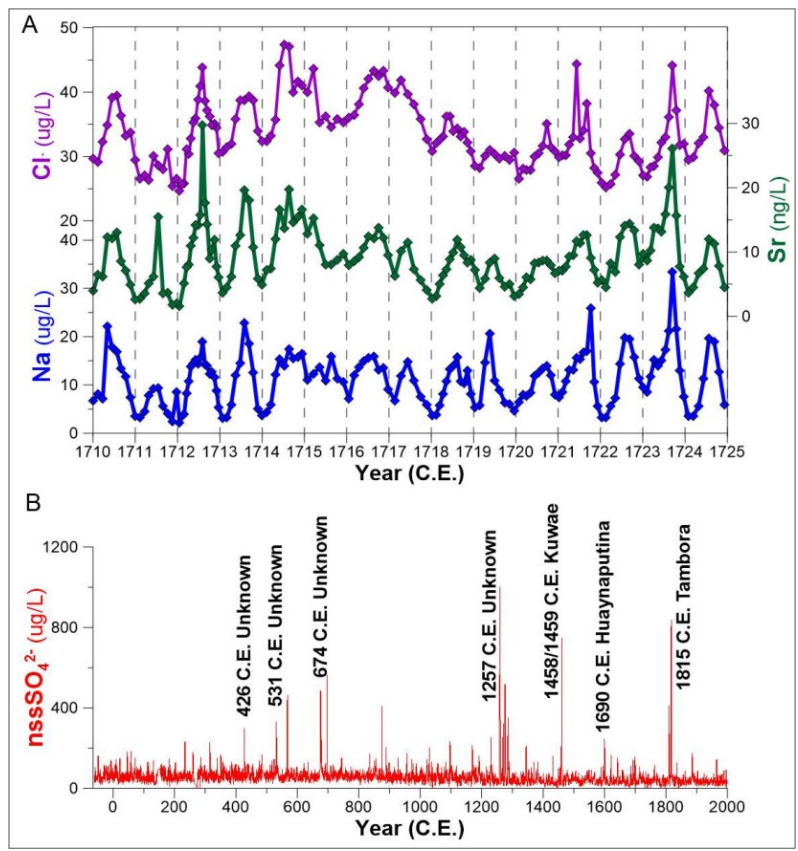

**Figure S2.** SPRESSO ice core timescale development using a CCI software package [*Kurbatov et al.*, 2005]: A) example of annual variations in Na (ug/L), Sr (ng/L) and Cl⁻ (ug/L) concentrations; B) nssSO$_4^{2-}$ record (ug/L) showing peaks attributed to major tropical volcanic eruptions.

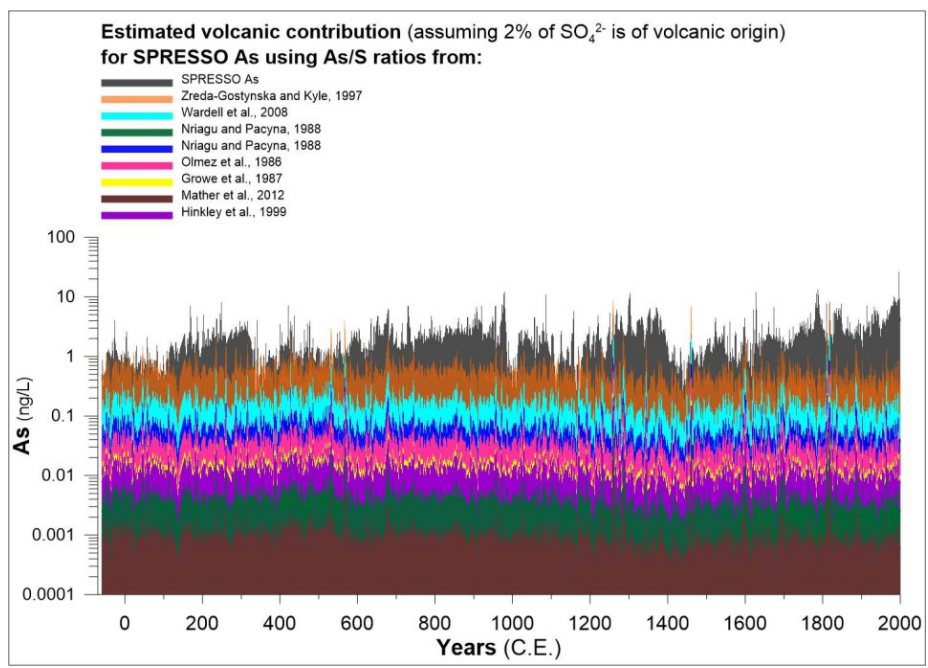

**Figure S3.** Estimated volcanic contribution for As using As/S values reported from volcanic plumes [*Olmez et al.*, 1986; *Crowe et al.*, 1987; *Nriagu and Pacyna*, 1988; *Zreda-Gostynska et al.*, 1997; *Hinkley et al.*, 1999; *Wardell et al.*, 2008; *Mather et al.*, 2012].

**Table S1.** Average method blank concentration (blank), method detection limit (MDL), minimum, maximum and mean sample concentration SPRESSO ice core, for elements used in this study. All concentrations are in ng/L.

| Element | Blank | MDL | SPRESSO concentrations | | |
|---|---|---|---|---|---|
| | | | Mean | Min | Max |
| As$^{75}$(LR) | 0.19 | 0.13 | 1.82 | <MDL | 26.15 |
| La$^{139}$(LR) | 0.02 | 0.001 | 0.37 | <MDL | 30.36 |
| Ce$^{140}$(LR) | 0.01 | 0.001 | 0.77 | 0.004 | 62.26 |
| Pr$^{141}$(LR) | 0.002 | 0.001 | 0.09 | <MDL | 7.08 |
| Na$^{23}$(MR) | 303 | 0.01 | 12187 | 0.029 | 413509 |
| Ti$^{47}$(MR) | 1.62 | 3.01 | 35.48 | <MDL | 2062.02 |
| SO$_4^{2-}$ | 3090 | 750 | 57600 | 1180 | 1002890 |

MDL is defined as three times the standard deviation of 7 MilliQ (>18.2 MΩ) deonized water blanks passed through the entire melter system

LR denotes low-resolution ICP-SMS mode (m/Δm=300), and MR denotes medium-resolution mode (m/Δm=4000)